\documentclass[prl,twocolumn,notitlepage,showpacs,preprintnumbers,amsmath,amssymb,nofootinbib,APS,10pt,superscriptaddress]{revtex4-2}

\usepackage{dcolumn}
\usepackage{bm}
\usepackage[spanish,english]{babel}
\usepackage[utf8]{inputenc}
\usepackage{amsmath,amssymb,amsfonts,latexsym,cancel}
\usepackage{graphicx}
\usepackage{color}
\usepackage{soul}
\usepackage{ulem}
\usepackage{hyperref}
\usepackage{slashed}
\usepackage{float}
\usepackage{xcolor}

\allowdisplaybreaks

\newcommand{\be}{\begin{equation}}
\newcommand{\ee}{\end{equation}}
\newcommand{\bea}{\begin{eqnarray}}
\newcommand{\eea}{\end{eqnarray}}

\begin{document}

\title{Comment on ``Gravitational Pair Production and Black Hole Evaporation''}

\author{Antonio Ferreiro}

\affiliation{Institute for Mathematics, Astrophysics and Particle Physics, Radboud University, 6525 AJ
Nijmegen, The Netherlands}
\email{antonio.ferreiro@ru.nl}

\author{Jos\'e Navarro-Salas} 

\affiliation{Departamento de Fisica Teorica and IFIC, Centro Mixto Universidad de Valencia-CSIC, 
Burjassot-46100, Valencia, Spain}

\email{jnavarro@ific.uv.es}
\author{Silvia Pla}  

\affiliation{Theoretical Particle Physics and Cosmology, King's College London, WC2R 2LS, UK}

\email{silvia.pla\_garcia@kcl.ac.uk}

\date{\today}
\maketitle

In the Letter \cite{paper}, the authors claim to explore a new avenue to address black hole  radiation within the general framework of quantum field theory interacting with a prescribed external background. This is a well-established field and one of its main predictions is the spontaneous creation of particles. 
The first fundamental example is the creation of pairs by a constant electric field \cite{Schwinger51}. 
The production rate is derived from the imaginary part of the ``effective action'' $W$, where $\langle 0_+|0_-\rangle = e^{iW}$ is the vacuum persistence amplitude. 
Gravitational pair creation was first derived in a cosmological scenario \cite{Parker69}, where the frequency-mixing mechanism was introduced as the new basic tool 
 \cite{Birrell-Davies, Parker-Toms}. The  frequency-mixing mechanism fully reproduce the Schwinger effect \cite{NN, GMM94}, showing the self-consistency of the general framework. The last fundamental example 
is the Hawking effect. The late-time thermal radiation predicted by the frequency-mixing mechanism requires a time-dependent process (gravitational collapse) \cite{HawkingCMP}. During the collapse a transient and tiny contribution to particle creation could be expected. However, at late times, only the steady thermal radiation remains due to the  event horizon. 
Within the framework of quantum field theory in curved spacetime \cite{Birrell-Davies, Parker-Toms, Wald}, all derivations of particle production agree with Hawking's result. 
Only the introduction of new ingredients such as  backreaction effects or quantum gravity  may lead to deviations.     

However, the approach of \cite{paper} is still in the realm of this framework. Hence, it raises questions regarding the authors' conclusions, which appear to deviate from Hawking's findings.  
In the rest of 
this comment we will point out  physical inconsistencies of the main formula of \cite{paper}     
\bea \label{eq:W}
\operatorname{Im}(W)&=& \frac{\hbar N}{64 \pi} \int \mathrm{d}^4 x \sqrt{-g}\Big[\frac{1}{2}\left(\xi-\tfrac{1}{6}\right)^2 R^2 \label{eq:W}\\
&+&\frac{1}{180}\left(R_{\mu\nu\rho\sigma} R^{\mu\nu\rho\sigma}-R_{\mu\nu} R^{\mu \nu}\right)+\frac{1}{12} \Omega_{\mu \nu} \Omega^{\mu \nu}\Big]\ , \nonumber \\
&+& higher-order \ terms \nonumber
\eea
which is assumed to be the imaginary part of the effective action $-$obtained in the ``weak field approximation'' for massless fields$-$ 
and the origin of the predicted local pair production. 

 First, we stress that (\ref{eq:W}) is obtained to lowest order in a perturbative expansion, while  the standard way to obtain the non-perturbative Schwinger effect 
 using the ``weak field approximation" is to perform a {\it resummation of all terms} of the proper-time expansion for constant backgrounds \cite{constant,NP2021}. This infinite sum gives the Euler-Heisenberg action, from which the Schwinger pair production rates  can be derived \cite{dunne}. In  contrast, the authors claim to recover the  Schwinger effect just from the second order of the proper-time series $\mathcal{O}(s^2)$.  The  perturbative formula (\ref{eq:W}) predicts results contradicting, in general, the Schwinger pair creation rates. This tension  can be easily shown  
 for a constant electromagnetic background such that $|\vec E|=|\vec B|$ ($\Omega_{\mu \nu} \Omega^{\mu \nu}=0$). \eqref{eq:W} gives zero particle creation, 
irrespective of the value of the second electromagnetic invariant.  However, {\it if $\vec E$ and $\vec B$ are parallel} this result is in contradiction with the non-perturbative solution  \cite{Popov, dunne} 
 \be \textrm{Im}\, L_{{\textrm{eff}}}= \frac{q^2E^2}{16\hbar \pi^2} \sum_{n=1}^\infty \frac{(-)^{n+1}}{n \sinh n\pi}\,.\ee
For fermions, this non-vanishing result is enforced by the axial anomaly. 
Even more, it is well-known that the particle production is {\it exactly zero for purely magnetic configurations} \cite{Schwinger51,itzykson}, in sharp tension 
with the result 
derived from (\ref{eq:W}), which predicts a non-unitary outcome ($\operatorname{Im}(W) <0$ and $|\langle 0_+|0_-\rangle|^2 > 1$). 

The 
inconsistencies of (\ref{eq:W}) also extend to gravitational pair creation. Consider a {\it conformally invariant} neutral scalar field ($\xi=1/6$) in a FLRW 
space-time. It is a well-established exact result that {\it there is no particle creation} \cite{Parker69, Birrell-Davies, Parker-Toms}.
The same is true for any conformally invariant field theory (Maxwell theory or massless spin-$1/2$ fields). 
However, for this background configuration the second term in (\ref{eq:W}) is non-zero ($R_{\mu \nu \rho \sigma} R^{\mu \nu \rho \sigma}-R_{\mu \nu} R^{\mu \nu}\neq0$), and according to 
\cite{paper} it would lead to particle creation, which is not consistent with the mentioned exact result. 
This term is just the gravitational conformal anomaly, and in a FLRW space-time 
it only accounts for vacuum polarization 
\cite{Parker-Toms, Parker2}.

The above discussion shows the incompleteness of the imaginary part of the (lowest-order) effective action (\ref{eq:W}) used to derive pair creation. Its perturbative predictions are significantly different from well-known exact results. It is very unlikely that finite next-to-leading order terms correct the 
predictions of the lowest order term (\ref{eq:W}). Note that, according to \cite{paper} (supplementary material), these higher-order contributions to (\ref{eq:W}) 
are inverse power of the mass, such that the massless limit may even be problematic. A more careful analysis including e.g., non-local terms could improve the predictions for particle creation consistently with well-established results. 

In summary, although the 
application to 
a Schwarzschild spacetime 
might appear appealing, the 
inconsistencies found in electrodynamics and cosmology raises  serious doubts 
regarding its main claim.

{\it Acknowledgements.} We thank C. Schubert for very useful comments. This work is supported by the Spanish Grants PID2020-116567GBC2-1  and PROMETEO/2020/079 (Generalitat Valenciana).  A.F.~is supported by the Margarita Salas fellowship MS21-085 of the University of Valencia. S.P. is supported by the Leverhulme Trust, Grant No. RPG-2021-299.

\section{Appendix}

In this appendix we give  additional details of  some aspects of our Comment not included in the main text. 

The authors of \cite{paper} 
argue in \cite {reply} that some of the  inconsistencies pointed out in the main text could be solved as a side effect of the higher order corrections  in (\ref{eq:W}). 
However, higher-order terms are not given, and it is only mentioned that the next-order term should be quartic in curvature.
We now show with more detail that this proposed way out of the  criticism does not work. 

It is very unclear what kind of local higher-order curvature terms could be added for a massless field to restore consistency with the exact result 
for parallel electric and magnetic fields \cite{Popov, dunne}
\be \label{exact}
\textrm{Im}\, L_{{\textrm{eff}}}= \frac{q^2EB}{16\hbar \pi^2} \sum_{n=1}^\infty \frac{(-)^{n+1}}{n \sinh (\frac{B}{E}n\pi)}\,.\ee
The prediction derived from (\ref{eq:W}) is (we take a single charged massless scalar field, $N=1$) 

\be \label{0LetterEB}\textrm{Im}\, L_{{\textrm{eff}}}^{Letter}= \frac{q^2 (E^2-B^2)}{192 \hbar \pi} + \ corrections\ . \ee
For $B\ll E$ the potential corrections should be negligible and we find agreement 
\be \label{accident}\textrm{Im}\, L_{{\textrm{eff}}}= \textrm{Im}\, L_{{\textrm{eff}}}^{Letter}= \frac{q^2 E^2}{192 \hbar\pi}  \ . \ee
For $E\ll B$, however, the potential corrections are negligible in any case and a major disagreement is found 
\be \textrm{Im}\, L_{{\textrm{eff}}}=0 \neq L_{{\textrm{eff}}}^{Letter}= -\frac{q^2 B^2}{16 \pi^3} <0\ . \ee
The negative result is incompatible with unitarity. 
For $E=B$ the potential higher-order corrections cannot correct the sharp discrepancy found at the leading order.  Note that the Schwinger formula (\ref{exact}) is proportional to $EB$, and not to $(EB)^2$. 
This means that the next-to-leading order correction 
will never capture the Schwinger effect for this configuration.


In the main text we also point out a tension between a well-established result in cosmology (i.e., the absence of particle creation for a massless and conformally coupled scalar field in a  FLRW spacetime) and the prediction obtained from the quadratic gravitational terms in (\ref{eq:W}). In \cite{reply}, the authors emphasize the extra disclaimer that the formula (\ref{eq:W}) should only be applied to stationary  configurations. However, it is easy to find a stationary counterexample within the FLRW spacetimes: de Sitter space. The application of (\ref{eq:W}) for a single real massless scalar field with $\xi=1/6$ predicts a non-unitary particle production result
\be \label{0LetterdeSitter}\textrm{Im}\, L_{{\textrm{eff}}}^{Letter}=  \frac{-12 \hbar H^4}{192\pi} < 0 \ . \ee 

Furthermore, 
we also want to discuss two important issues, not included in the main discussion due to the lack of space.



 We have pointed out that the leading term in (\ref{eq:W}) is proportional to the integral of the trace anomaly (this was previously noted in \cite{Chernodub} and also confirmed in \cite{reply}). However it is very difficult to attribute the particle creation phenomena to the trace anomaly, and hence to the leading terms in (\ref{eq:W}). For several reasons:

i) The trace anomaly by itself is independent of the quantum state. To obtain the late-time Hawking radiation for a stationary 4D Schwarzschild black hole, one must assume the Unruh state. Under the assumption of a Boulware state for a compact object, the anomalous trace is the same, but the geometry is radiation-free 
\cite{Candelas, Fabbri-Navarro}. 
[We also note  in passing that the Unruh vacuum breaks the time-reversal symmetry and generates a thermal flux $\langle U| T^t_r|U\rangle$ at infinity $r\to \infty$ at the Hawking temperature. This outgoing positive flux is exactly counterbalanced by a similar incoming negative flux through  the future event horizon \cite{Candelas, Fabbri-Navarro}. 
This energy-conserving mechanism  is missing from the heuristic description of gravitational pair creation given in \cite{paper}.
\footnote{We thank R. Emparan for suggesting  this point.}]

ii) There is no trace anomaly in odd dimensions, while the temperature of stationary black holes is, in general, non-zero \cite{Kanti, Emparan}.  

iii) The electromagnetic trace anomaly in two-dimensions is zero. However, there is a non-trivial Schwinger effect \cite{Peskin}.


Finally, we would like to stress that the perturbative formula (\ref{eq:W}) was previously obtained within the conventional perturbative expansion in powers of the electric coupling constant and also in a perturbative expansion around flat space. Those results are given,   for the massive and massless fields, in \cite{itzykson} and  \cite{Frieman, Verdaguer}, respectively  \footnote{We thank F.D. Mazziteli for the reference \cite{Frieman}.}. The formulas obtained in those references generalize expression (\ref{eq:W}).  For example, for the electromagnetic case (Minkowski space) one has (see \cite{itzykson} Section 4-3)
\bea \label{eq:IZformula}
\textrm{Im}(W)&=&\frac{\alpha}{12} \int d^4 q \,\theta(q^2-4 m^2)\left[|\mathbf{E}(q)|^2-|\mathbf{B}(q)|^2\right] \nonumber \\ &\times&(1-\frac{4 m^2}{q^2})^{3 / 2}
\eea
It is easy to check that the effective Lagrangian for a pure magnetic field is exactly zero (neither negative nor positive), improving this way (\ref{eq:W}). For a  constant electric field the perturbative result \eqref{eq:IZformula} disagree, as expected, with the exact result. The  perturbative expansion will never capture the (non-perturbative) formula of the Schwinger effect, as it is well explained in the refereed textbook. \\

All the above remarks  reinforce our central claim  in the main text of the Comment: (\ref{eq:W}) is an incomplete expression to account for particle creation in a way that is consistent with well-established exact results. \\

\end{document}